# THERMALLY ASSISTED FLUX FLOW IN MgB$_2$ : STRONG MAGNETIC FIELD DEPENDENCE OF THE ACTIVATION ENERGY

A. Sidorenko[a,d,e], V. Zdravkov[a,b], V. Ryazanov[b], S. Horn[c], S. Klimm[c], R. Tidecks[c],

A. Wixforth[c], Th. Koch[d], Th. Schimmel[d,e]

[a]Institute of Applied Physics, MD-2028 Kishinev, Moldova
[b]Institute of Solid State Physics, RU-142432 Chernogolovka, Russia
[c]Institut für Physik, Universität Augsburg, D-86159 Augsburg, Germany
[d]Institute of Nanotechnology (INT), Forschungszentrum Karlsruhe, D-76021 Karlsruhe, Germany
[e]Institute of Applied Physics, Universität Karlsruhe, D-76128 Karlsruhe, Germany

The origin of the resistive transition broadening for MgB$_2$ thin films was investigated. Thermally activated flux flow is found to be responsible for the resistivity contribution in the vicinity of $T_c$. The origin of the observed extraordinary strong magnetic field dependence of the activation energy of the flux motion is discussed.

## 1. Introduction

The discovery of superconductivity in MgB$_2$, a material with a hexagonal layered crystal structure and the highest critical temperature, $T_c$ = 39 K, found for an intermetallic superconducting compound [1], raised questions about its transport properties. This strong type-II superconductor with a large Ginzburg-Landau parameter $\kappa \approx 26$, a magnetic penetration length $\lambda(0)$ = 140-180 nm [2] and short coherence lengths, $\xi_c(0)$ = 2.3 nm, $\xi_{ab}(0)$ = 6.8 nm [3] has a rather high critical current density up to $j_c \sim 1.6 \times 10^7$ A/cm$^2$ at 15 K [4]. The latter finding makes the novel superconductor very attractive for technical applications. On the other hand, a broadening of the superconducting transition, as found in resistivity measurements, would severely limit the applicability of MgB$_2$. Therefore, it is important to study the mechanism which causes this broadening.

Superconducting transition broadening in the presence of a magnetic field can have different reasons. It may be caused by an inhomogeneous microstructure of polycrystalline samples with additional phases having different $T_c$. Moreover, fluctuations play an important role in the vicinity of the superconducting transition especially for low-dimensional and layered superconductors with a short coherence length and a high $T_c$, as for $MgB_2$. Finally, thermally activated dissipation due to vortex lines motion yields a broadening of the transition.

The first reason of superconducting transition broadening can be avoided by improving the technological process of sample preparation. The second two mechanisms are of fundamental nature and, therefore, attracted more attention from experimentalists as well as from theorists. The fluctuation governed superconducting transition broadening was investigated for high quality homogeneous $MgB_2$ films [5] and single crystals [6].

A lack of information up to now, however, exists concerning the broadening of the resistive transition due to thermally activated flux creep or flux flow (TAFF) processes below the critical temperature for $MgB_2$. In a type-II superconductor in the mixed state the flux lines are fixed at "pinning centers", i.e. for example at defects or impurities. The main mechanism of the flux creep, yielding the resistive transition broadening in a magnetic field, is the thermal activation the flux-line motion over the energy barrier, $U_0$, of the pinning center [7].

The layered structure of $MgB_2$ is expected to influence the magnetic flux penetration and motion leading to a resistive transition broadening similar to the case of high-$T_c$ superconductors [7, 8] and artificially multilayered systems [9,10]. On the other hand magnesium diboride exhibits an exceptional magnetic behavior with dendritic magnetic instabilities for vortex penetration [11] and "noise-like" jumps of the magnetization in an applied magnetic field [12], which should influence the resistive behavior of this novel superconducting material.

The present work deals with an experimental investigation of the resistive transition broadening of magnesium diboride thin films, caused by the TAFF mechanism.

2. **Experimental**
2.1. **Sample preparation**

The $MgB_2$ films with the thickness 100-1000 nm were fabricated by DC-magnetron sputtering from $Mg$-$MgB_2$ composite targets (target diameter 32 mm, thickness 5 mm) prepared by a hot-pressing procedure from 99.9 % purity Mg powder and 98 % purity $MgB_2$ powder (Alfa Aesar). The films were deposited on (100)-oriented sapphire substrates, and on ($128^0$ rot)-$LiNbO_3$ substrates. During the sputtering process the substrate temperature was kept at approx. $100^0$C. The deposition rate was 1.3 nm/s. Next, the deposited film was annealed *ex situ* at $850^0$C in a Mg vapor atmosphere. Further details of the sample preparation are described in [5]. An X-ray diffraction study revealed a textured (101)-oriented structure of the films deposited on sapphire substrates, and policrystalline films on $LiNbO_3$. From the deposited films samples were cut as 1.5 mm wide strips using a diamond cutter. Platinum wires with 50 μm diameter were attached by silver paste for four-probe resistance measurements.

**2.2. SEM and AFM characterization of the samples**

A Gemini-982 system from Leo was used for the Scanning Electron Microscopy (SEM). The analysis of the SEM images was made with Leo software. For the Atomic Force Microscopy (AFM), Nanoscope-III/IIIa driven Multimode Systems equipped with 130 μm vertical engage scanners and with commercial and homebuilt phase imaging units were used. The experiments were made using type CSC21 (contact mode, force load approx. 5 nN) and type NSC15 (intermittent contact mode, set point 0.75) silicon cantilevers from NT/MDT. All AFM measurements were made under ambient conditions. The statistical analysis of the AFM-images was made with SPIP software from Image-Met.

SEM shows a rather rough morphology of the films on $LiNbO_3$ substrate having a trigonal crystal structure, not matching to the hexagonal $MgB_2$ structure. Fig.1 presents an SEM image of a $MgB_2$ film on $LiNbO_3$ with grain size variation of a wide range, from 0.2 μm up to 6 μm. The surface of the film on a $LiNbO_3$ substrate is so rough that the topography could not be resolved completely with AFM because of scan range limitations of the instrument. (i.e. the height of some grains of the film was more than 5 μm).

The SEM investigations of $MgB_2$ films on single crystalline (100) sapphire substrate (not shown here) demonstrated in comparison to the films on $LiNbO_3$ a much more homogeneous microstructure with a very smooth surface. These smooth films were investigated by AFM to get detailed information about roughness and material contrast.

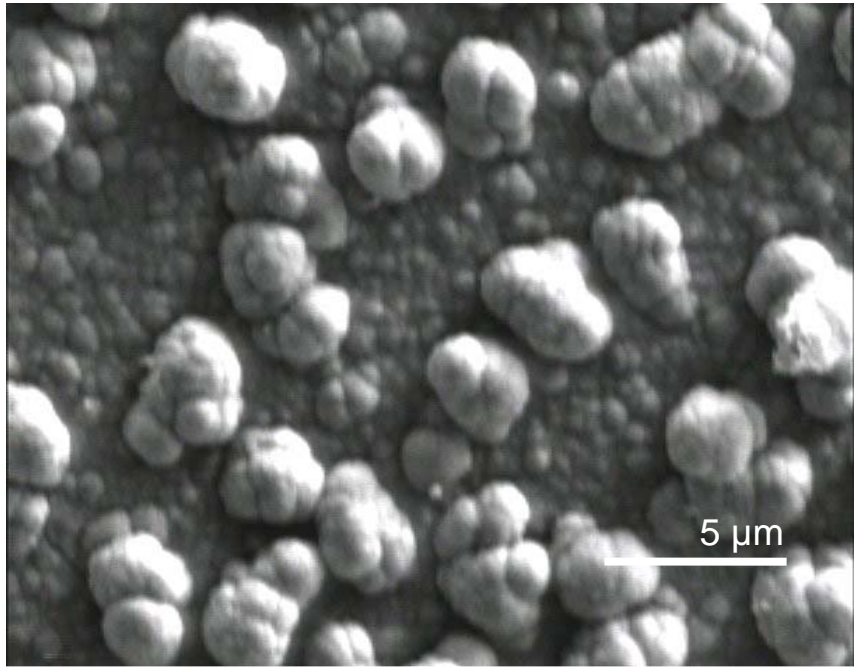

Fig.1 Electron microscope (SEM) image of 5μm thick $MgB_2$ film on a $LiNbO_3$ substrate (scan size 20 μm x 15 μm, 5KV); a rough island covered surface with a wide distribution of island size up to 5 μm were imaged.

The AFM images of the surface for a 400 nm thick $MgB_2$ film on sapphire substrate (Fig. 2) were taken at several different sample positions both in the contact mode and tapping mode. The surface is homogenously covered with flat islands (average height: 13 nm) with an average diameter of 50 nm (Fig. 2c, Fig. 2d). The results show this planar structure even for big scan sizes (Fig. 2a). Uniformly distributed island groups (Fig. 2a, Fig. 2b) with diameters between 100 nm and 3 μm, a height of about 100 nm and a typical group to group distance of about 12 μm (Fig. 2a) are located on the surface of the film. These groups also consist of islands of an average size of 50 nm. The area coverage of the sample surface with these island groups is approximately 11%. Tapping Mode Phase Imaging [13] and Friction Force Microscopy [14] did not show any significant material contrast on the film surface. The absence of differences in Phase Imaging shows that the values for adhesion and elasticity stay the same all over the observed surface areas. In addition the results of the Friction Force Microscopy show a constant material contrast in these areas, too. This accentuates the homogeneity of the investigated sample surfaces. The RMS roughness values of the $MgB_2$ films on sapphire substrate are between 6.3 nm and 8.2 nm. This fact, together with the absence of the material contrast demonstrates the smooth and homogeneous character of the film.

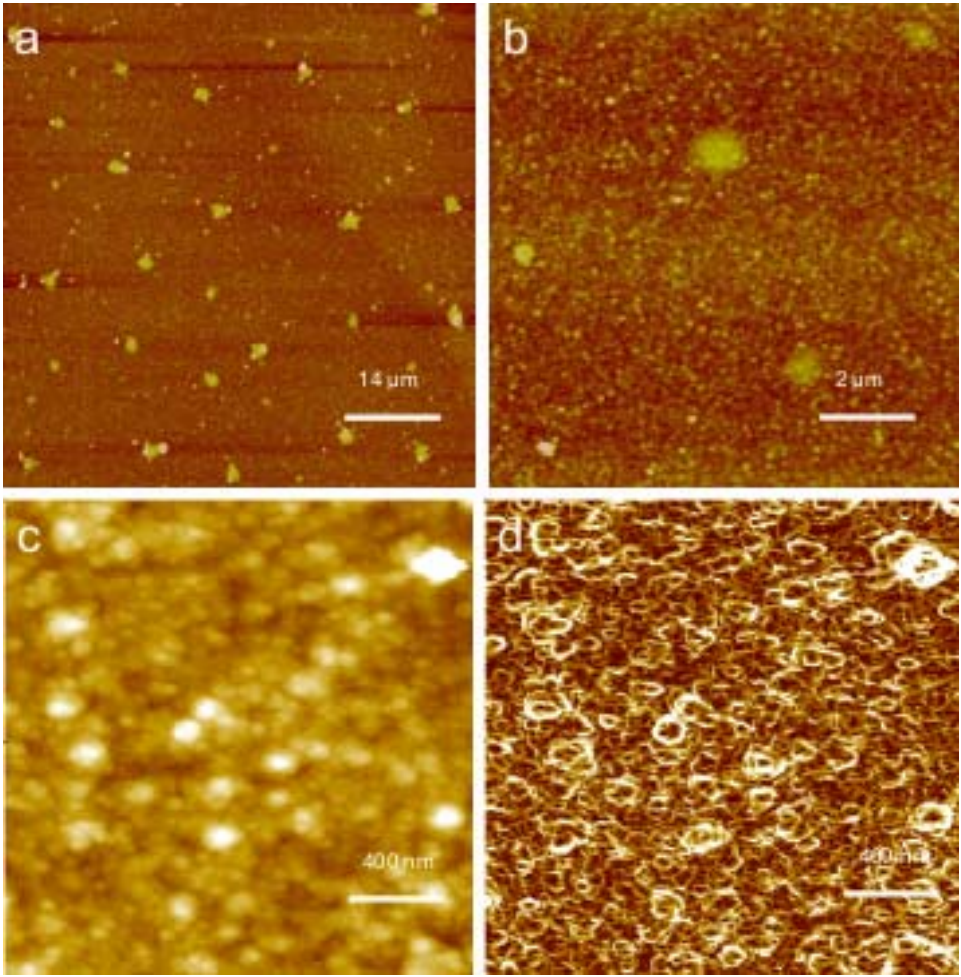

Fig.2. Atomic Force Microscopy images taken from the surface of a 400 nm thick $MgB_2$ film on (100) sapphire substrate: (a) tapping mode, scan size: 68 μm x 68 μm, z-scale 100 nm, RMS roughness 8.2 nm +/- 1.5 nm; (b) contact mode, scan size: 10 μm x 10 μm, z-scale 100 nm, RMS roughness 6.3 nm +/- 1.2 nm; (c) contact mode, scan size: 2 μm x 2 μm; z-scale 100 nm, RMS roughness 6.3 nm +/- 1.2 nm; (d) derivative of c, z-scale 82,6º; The surface consists homogenously of small islands with an average diameter of 49 nm +/- 15 nm (c, d). In addition uniformly distributed island groups with a size up to 3 μm can be seen (a, b). These groups consist of densely packed islands with an average size of 50 nm. No significant material contrast was found in Tapping Mode Phase Imaging and Contact Mode Friction Force images.

## 2.3. Resistance measurements

The resistance measurements, $\rho(T)$, of the MgB$_2$ samples were performed by a conventional four-probe method using an AC resistance bridge (Linear Research, LR700) in a $^4$He cryostat (Oxford Instruments) equipped with a 12 Tesla superconducting solenoid. The temperature $T$ was measured with a carbon-glass thermometer with an accuracy of 1-5 mK. The critical temperature $T_c$ was determined from the midpoints of the $\rho(T)|_{B=const}$ curves.

## 3. Results and Discussion

Fig. 3 shows the resistive transitions $\rho(T)$ at several magnetic fields, $B$, perpendicular to the MgB$_2$ film plane for one of the investigated samples. The transition width is about 0.3 K in zero and low magnetic fields but increases up to ~2 K for high fields. Usually, the broadening of the lower parts of the resistive transition, $\rho(T) < 1\%\,\rho_n$ (where $\rho_n$ is the resistivity in the normal state just above the transition), in a magnetic field for layered superconductors is interpreted in terms of a dissipation of energy caused by the motion of vortices [8]. This interpretation is based on the fact that for the low-resistance region, the resistance is caused by the creep of vortices so that the $\rho(T)$ dependences are of the thermally activated type described by the equation

$$\rho(T,B) = \rho_0 \exp[-U_0/k_B T] \qquad (1)$$

Here, $U_0$ is the flux-flow activation energy, which can be obtained from the slope of the linear parts of an Arrhenius plot (according to Eq.(1)) and $\rho_0$ is a field-independent pre-exponential factor. Investigations of high-$T_c$ superconductors and artificial multilayers showed that the activation energy exhibits different power-law dependences on a magnetic field, i.e. $U_0(B) \sim B^{-n}$ [7,8,9]. Since Eq.(1) with a temperature-independent $U_0$ is used to describe our experiments, the values of $U_0$ should be deduced only from the limited temperature intervals below $T_c$, in which the data of the Arrhenius plot of $\rho(T)$ yield straight lines.

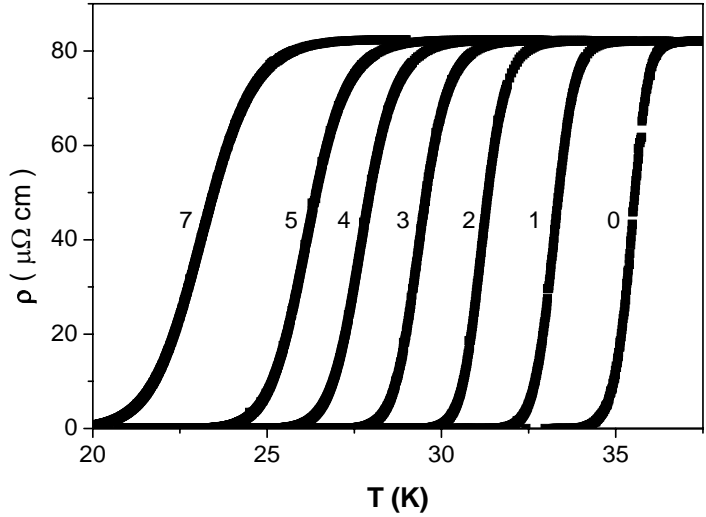

Fig. 3 Resistive transitions $\rho(T)$ for a 400 nm thick $MgB_2$ film at different values of the magnetic field perpendicular to the film plane: curves 0 to 7 correspond to B = 0, 1, 2, 3, 4, 5, 7 Tesla.

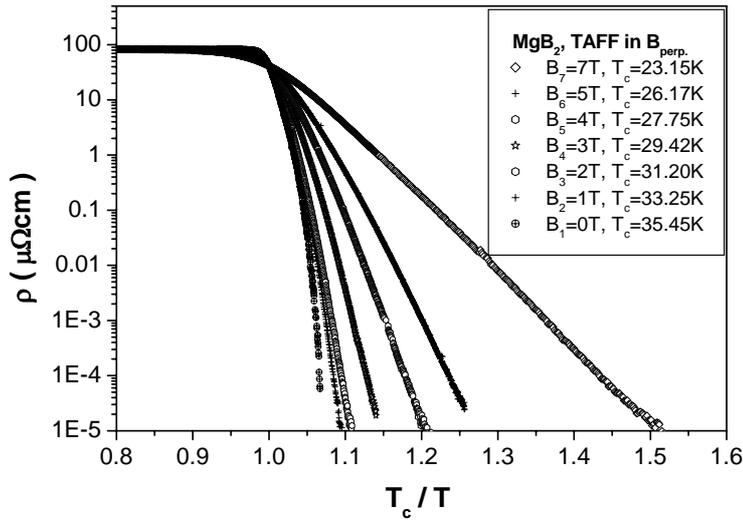

Fig. 4 Arrhenius plot of $\rho(T)|_{B=const}$ for the sample, presented in Fig.3. From the slope of the linear parts of the curves the values of the activation energy $U_0$ are obtained.
The Arrhenius plot in Fig. 4 presents the data of Fig. 3 as $ln(\rho)$ against $(T_c/T)$.

The straight-line behavior over five orders of magnitude of the resistance indicates that the resistive behavior of the $MgB_2$ film is caused by the TAFF-process as described by the Arrhenius law given in Eq. (1). The best fit of the experimental data $\rho(T)|_{B=const}$ by Eq. (1) yields values of the activation energy, ranging from $U_0/k_B$ = 10000 K in low magnetic field down to 300 K in the high field region, as shown in Fig. 5. Compared to the power law $U_0(B) \sim B^{-n}$ with the exponent $n \leq 1$, usually observed for other layered systems [8-10,15,16], $MgB_2$ shows a much stronger field dependence of the activation energy in the high magnetic field region ($B > 1T$). This is clearly demonstrated by Fig. 5, where our data for $MgB_2$ are shown together with a typical $U_0(B)$ dependence for a high-$T_c$ superconductor (data for a Bi-Sr-Ca-Cu-O sample taken from Palstra et al. [15]).

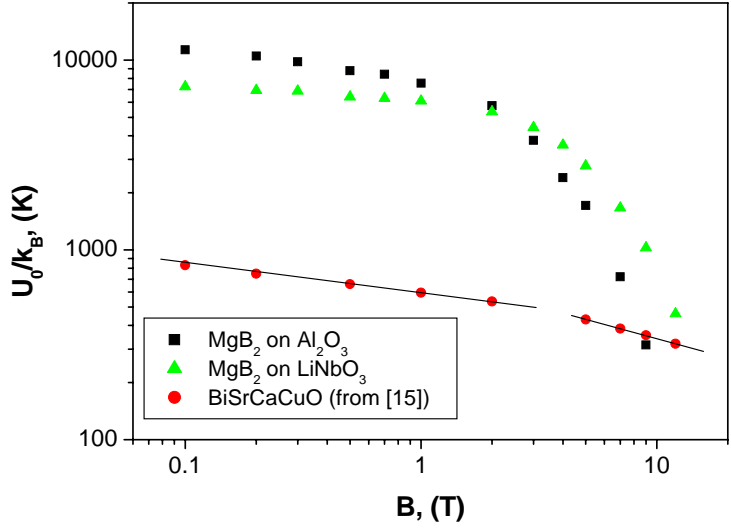

Fig.5. Dependence of the activation energy $U_0/k_B$ on magnetic field for the investigated samples (solid quadrates, 400 nm thick $MgB_2$ film deposited on sapphire; solid triangles, 5μm thick film deposited on $LiNbO_3$). For comparison, the weak power-law dependence $U_0 \sim B^{-n}$, n = 1/6 and n = 1/3 for two linear parts of the $U_0(B)$ dependence, for a high-$T_c$ superconductor (solid circles, data for Bi-Sr-Ca-Cu-O sample taken from [15]) is given.

The rapid decrease of the activation energy for $MgB_2$ in field region $B > 1T$ reflects a dramatic loss of the current carrying capabilities of the superconductor due to the weakening of the flux-line pinning with increasing magnetic field. A possible reason for the unusually strong magnetic field dependence of the activation energy of the TAFF process in $MgB_2$, observed in the present work, may be the appearance of thermo-magnetic instabilities, considered by Mints and Rakhmanov [17], leading to complex flux dynamics, such as the dendritic flux instability in $MgB_2$ films, found recently [11] for c-axis textured films in a magnetic field perpendicular to the film plane. The magneto-optical measurements demonstrate a "fractal-like" structure of the flux penetration with a strongly increasing amount of the flux-dendrite density with increasing magnetic field. Mesoscopic flux jumps appear as a result of the thermo magnetic instability [18]. More generally, thermal instabilities, analyzed by M. Tinkham [19], lead to disastrous consequences for superconducting magnets and cables because the material may rapidly heat up due to the dissipation of energy associated with flux creep. Therefore for thermal stability of a superconductor, an enhancement of the temperature must yield an increased outflow of heat to the surrounding material which has to correspond to the increased dissipation due to more rapid flux motion.

As one can conclude from the reported experimental results the strong magnetic field dependence of the activation energy $U_0(B)$ may be an intrinsic property of $MgB_2$ as a quite similar strong $U_0(B)$ dependence was observed both for samples with a homogeneous smooth microstructure prepared on sapphire substrate and for very rough films deposited on $LiNbO_3$ substrate.

The unusual flux creep behavior of magnesium diboride needs further investigation, especially in view of future applications, for which an increase of the flux line pinning and a thermal stabilization of wires and tapes are necessary for electrical transport with high current density.

**Acknowledgments**

The authors are grateful to P. Ziemann, J. Eisenmenger, E.-H. Brandt, A. Zaikin for useful discussions, to A. Rossolenko and O. Kroemer for experimental assistance, to the A.v.Humboldt Foundation for the donation of "Coolpower-4.2GM" and "PLN-106" refrigerators. This work was partially supported by BMBF, Project Nr. MDA02/002, by the Deutsche Forschungsgemeinschaft within the DFG-Center for Functional Nanostructures CFN and by INTAS; Project Nr. 03-55-1856.